%Paper: hep-th/9310112
%From: NAPPI@sns.ias.edu
%Date: 18 Oct 1993 13:31:25 -0400 (EDT)
%Date (revised): 19 Oct 1993 13:49:56 -0400 (EDT)

\def\IR{{\hbox{{\rm I}\kern-.2em\hbox{\rm R}}}}
\def\IB{{\hbox{{\rm I}\kern-.2em\hbox{\rm B}}}}
\def\IN{{\hbox{{\rm I}\kern-.2em\hbox{\rm N}}}}
\def\IC{{\ \hbox{{\rm I}\kern-.6em\hbox{\bf C}}}}

\def\IZ{{\hbox{{\rm Z}\kern-.4em\hbox{\rm Z}}}}

\def\underarrow#1{\vbox{\ialign{##\crcr$\hfil\displaystyle
{#1}\hfil$\crcr\noalign{\kern1pt
\nointerlineskip}$\longrightarrow$\crcr}}}
% use of underarrow
%A~~~\underarrow{a}~~~B
%
\def\d{{\rm d}}

\input phyzzx
\overfullrule=0pt
\tolerance=5000
\overfullrule=0pt
\twelvepoint

\twelvepoint
\date{}
\line{ \hfill hep-th/9310112}
\line {\hfill IASSNS-HEP-93/61}
\line{\hfill September 1993}
\titlepage
\title{A  WZW model based  on a non-semi-simple group}
\vglue-.25in
\author{Chiara R. Nappi\foot{Research supported in part by
the Ambrose Monell Foundation.} and Edward Witten
\foot{Research supported in part by NSF Grant
PHY91-06210.}}
\medskip
\address{School of Natural Sciences
\break Institute for Advanced Study
\break Olden Lane
\break Princeton, NJ 08540}
\bigskip
\abstract{We present a conformal field theory which desribes a
homogeneous four dimensional Lorentz-signature space-time.
The model is an ungauged
WZW model based on a central extension of the Poincar\'e algebra.
 The central charge of this
theory is exactly four,  just like four dimensional Minkowski space.
The model can be interpreted as a four dimensional monochromatic
plane wave.  As there
are three commuting
isometries, other interesting geometries are expected to
emerge via $O(3,3)$ duality.}
\endpage
\REF\Jackiw{D. Cangemi and R.Jackiw, Phys. Rev. Lett. {\bf 69} (1992)
233-236}
\REF\Witten{E. Witten, Comm. Math. Phys. {\bf 92}, 455-472 (1984)}
\REF\Nappi{C.R. Nappi, Phys. Rev. {\bf D21}, 418 (1980)}
\REF\Halpern{M.B. Halpern and E. Kiritsis, Mod. Phys. {\bf A4} (1989)
1373-1380 }
\REF\olive{D. Olive, E. Rabinovici, and A. Schwimmer, to be published;
also J. Distler, to appear.}
\REF\Rosen{H.W. Brinkman, Proc. Natl. Acad. Sci. (U.S.) {\bf 9}, 1
(1923); N. Rosen, Phys. Z. Sowjetunion {\bf 12}, 366 (1937)}

In this paper, we will describe a new example of a
homogeneous (but not isotropic) four dimensional space-time
-- with Lorentz signature -- that can be constructed
from an ungauged WZW model.  The model has $c$ (or in the
supersymmetric
case $\widehat c$) equal to four, so it can be directly substituted
for four dimensional Minkowski space and regarded as a solution of any
more or less realistic string theory.  It describes a special case
of a  monochromatic plane wave with more than the usual symmetry.

The model is a WZW model based on a certain non-semi-simple Lie
algebra
of dimension four.
The algebra has the following explicit description:
$$[J, P_i]=\epsilon_{ij}P_j \qquad [P_i, P_j]=\epsilon_{ij}T\qquad
[T,J]=[T,P_a]=0\eqn\zero$$
This algebra
 is a central extension of the 2D Poincar\'e algebra to which it
reduces if
one sets $T=0$.  We will call the corresponding simply-connected group $R$.
% From \zero\ it
%is obvious that the only non-zero ones in the present case
%are $f^4_{ij}=\epsilon_{ij}$
%and $f^j_{3i}=\epsilon_{ij}$, with $i,j=1,2$.

In general, given a Lie  algebra with generators
$T_A$ (here $T_A= P_1, P_2, J, T$), and structure constants $f^D_{AB}$
(so $[T_A,T_B]=f^D_{AB}T_D$), to
define a WZW model, one needs a bilinear form
$\Omega_{AB}$ in the
generators $T_A$, which is symmetric ($\Omega_{AB}=\Omega_{BA}$),
invariant
$$f^D_{AB}\Omega_{CD} + f^D_{AC}\Omega_{BD} = 0\eqn\inv$$
and non-degenerate (so that there is an inverse matrix $\Omega^{AB}$
obeying $\Omega^{AB}\Omega_{BC}=\delta^A_C$).
\foot{Usually, $\Omega$ must
obey a certain integrality condition, but that will have no analog
here because of the simple structure of $R$.  That will be clear from
the ability below to reduce the Wess-Zumino term to an integral over
$\Sigma$, without introducing any singularities.}
Then the WZW action on a Riemann surface $\Sigma$
is
$$S_{WZW}(g) = {1\over 4\pi}\int_\Sigma \d^2\sigma ~\Omega_{AB} A^A_\alpha
A^{B\alpha} + {i\over 12\pi}\int_B \d^3\sigma~
\epsilon_{\alpha\beta\gamma}A^{A\alpha}A^{B\beta}A^{C\gamma}
\Omega_{CD}f^D_{AB} \eqn\wzw$$
where the $A_\alpha^A$'s are defined via
 $ g^{-1}\partial_\alpha g = A^A_\alpha T_A$.
Here $B$ a three-manifold with
boundary
$\partial B=\Sigma$, and $g$ is a map of $\Sigma$ to $R$ (extended in
an arbitrary fashion to a map from $B$).
Usually for semisimple groups one can choose the bilinear form
 $\Omega_{AB}= f^D_{AC}f^C_{BD}$ (which is equivalent to $\Tr T_AT_A$
with
the trace taken in the adjoint representation). However, for
non-semi-simple groups this quadratic form is degenerate;
and indeed for the algebra \zero\ one gets
$$\Omega_{AB}=\pmatrix{0&0&0&0\cr
0&0&0&0\cr
0&0&-2&0\cr
0&0&0&0\cr} \eqn\deg  $$
Nevertheless, the $R$ Lie algebra
does have another non-degenerate bilinear form  [\Jackiw]
{\it i.e.}
$$\Omega_{AB}=\pmatrix{1&0&0&0\cr
0&1&0&0\cr
0&0&0&1\cr
0&0&1&0\cr} \eqn\ndeg  $$
It is easily shown that the most general invariant quadratic form on
this
Lie algebra is a linear combination of these:
$$\Omega_{AB}=\pmatrix{k&0&0&0\cr
0&k&0&0\cr
0&0&b'&k\cr
0&0&k&0\cr}  \eqn\formm  $$
By setting $b=b'/k$ we can write $\Omega_{AB}$ and its inverse as
$$\Omega_{AB}=k\pmatrix{1&0&0&0\cr
0&1&0&0\cr
0&0&b&1\cr
0&0&1&0\cr}\qquad \Omega^{-1}=\Omega^{AB}= {1\over k}\pmatrix{1&0&0&0\cr
0&1&0&0\cr
0&0&0&1\cr
0&0&1&-b\cr}  \eqn\form  $$
This metric on the Lie algebra has signature $+++-$, and that will
therefore be the signature of the space-time described by the
corresponding WZW model.

In order to write \wzw\ explicitly we need to find the $A_A$'s. To
this purpose we use the following parametrization of the group
manifold:
$$g= e^{\sum_ia_iP_i}e^{uJ+vT}   \eqn\group$$
(In fact by using the commutation relations to move all of the $J$'s
and
$T$'s to the right, any group element can be uniquely brought to this
form; this is somewhat like the process of normal-ordering of an
exponential of a sum of creation and annihilation operators.)
By using
$$e^{uJ}P_ie^{-uJ} = \cos u P_i +  \epsilon_{ij}\sin u P_j \eqn\co$$
and
$$\partial_\alpha e^H = \int_0^1dxe^{xH}\partial_\alpha H e^{(1-x)H}      $$
we compute
$$\partial_\alpha g =  e^{\sum_ia_iP_i}(\partial_\alpha a_iP_i +{1\over 2}
\epsilon_{jk}T\partial_\alpha a_ja_k)e^{uJ+vT} + g \partial_\alpha
(uJ + vT)  \eqn\part$$
By using \part\ and \co\ we find
$$g^{-1}\partial_\alpha g = (\cos u\partial_\alpha a_k + \sin u
\epsilon_{jk}\partial_\alpha a_j)P_k  + (\partial_\alpha u)J  +
(\partial_\alpha v + {1\over 2}\epsilon_{jk}\partial_\alpha a_j a_k)T
\eqn\as     $$
from which we can read off the $A^A$'s and compute
$$A^A_\alpha A^{B\alpha}\Omega_{AB}= \partial_\alpha
a_k\partial^{\alpha}a_k + 2 \partial_{\alpha}v\partial^{\alpha}u +
b\partial_\alpha\partial^\alpha u +
\epsilon_{jk} \partial^{\alpha}u\partial_\alpha a_j a_k\eqn\kin$$
$$\epsilon_{\alpha \beta \gamma} A^{A\alpha} A^{B\beta} A^{C\gamma}
\Omega_{CD}f^D_{AB} =2\epsilon_{\alpha \beta \gamma}\times
3\partial^{\alpha}[u\partial^\beta a_1\partial^\gamma a_2]\eqn\wzterm$$
Finally the Lagrangian looks like
$${k\over {4\pi}}\int d^2\sigma [\partial_\alpha a_k \partial_\alpha
a_k + 2\partial_\alpha v\partial^\alpha u +
b\partial_\alpha u\partial^\alpha u +
\epsilon_{jk}\partial^\alpha u\partial_\alpha a_j a_k  + i2\epsilon_{\beta
\gamma}u\partial^\beta a_1 \partial^\gamma a_2 ]\eqn\total$$

By
identifying this Lagrangian with the $\sigma$-model action of the
form
$$S = \int \d^2\sigma (G_{MN}\partial_\alpha X^M{\partial^\alpha} X^N
+iB_{MN}\epsilon_{\alpha\beta}\partial^\alpha X^M\partial^\beta X^N),
\eqn\six$$
where $X^M=(a_1, a_2, u, v)$
one can
read off the background space-time metric and antisymmetric tensor field.
The space-time geometry  is described by a Lorentz signature
 metric $G_{MN}$
$$G_{MN}=\pmatrix{1&0&{a_2\over 2}&0\cr
0&1&-{a_1\over 2}&0\cr
{a_2\over 2}&-{a_1\over 2}&b&1\cr
0&0&1&0\cr}\qquad G^{-1}=G^{MN}=\pmatrix{1&0&0&-{a_2\over 2}\cr
0&1&0&{a_1\over 2}\cr
0&0&0&1\cr
-{a_2\over 2}&{a_1\over 2}&1&{a_1^2\over 4}+{a_2^2\over 4}-b\cr}
 \eqn\metric  $$
and $B_{12}=u$, and of course the dilaton is constant because of
the homogeneity of the group manifold. (In gauged WZW
models, the non-constant dilaton emerges from integration over the
gauge fields). The corresponding
space-time metric is
$$ ds^2= da_kda_k + 2dudv +\epsilon_{jk}da_ja_kdu + bdu^2 \eqn\name$$

As a WZW model, this model should be conformally invariant.
To check this, we first look at the one loop beta function
equations
$$\eqalign{&R_{MN} - {1\over 4}H_{MN}^2 - D_M D_N\phi=0\cr
 & D^L H_{LMN} + D^L\phi
H_{LMN} = 0\cr & -R + {1\over {12}}H^2 + 2\Delta^2\phi +
(\Delta\phi)^2 + \Lambda =0\cr}\eqn\uffa$$
where $H_{LMN} = D_{[L}B_{MN] }$ and
$H^2_{MN}=H_{MPR}H_N{}^{PR}$, $H^2=H_{MPR}H^{MPR}$,
and $\Lambda=2(c-4)/3\alpha'$.
One quickly finds that the only non-zero component of $R_{MN}$ is
$R_{33}=1/2$; and as the only non-zero component of $B$ is
$B_{12}=u$,
the only non-zero component of $H$ is $H_{312}=1$.
As $G^{33}=0$, $R=H^2=0$.  The only non-zero component of $H^2_{MN}$
is $H^2_{33}=2$.
Putting these pieces together, one verifies equations \uffa\
with $\Lambda=0$ and $c=4$.

A more direct way to see that the one loop beta function equations are
satisfied is to actually compute the one loop diagrams that
contribute to the beta functions. First of all, the only one loop
diagram that one can draw is the one with two $u$'s in the external
lines, and $a_i$'s in the internal lines. This is because
there is no $uu$
propagator, as $G^{33}=0$.
Moreover, by
comparing the interaction terms in the Lagrangian one notices that
they give opposite contributions to this one-loop diagram.
Indeed the interaction term contained in the space-time metric piece of the
Lagrangian is
$$\epsilon_{jk}a_k\partial_\alpha a_j \partial^\alpha u =  \eta_{\beta \gamma}
\epsilon_{jk} a_k \partial^\beta a_j
\partial^\gamma u \eqn\coupl $$
where $ \eta_{\beta\gamma}$ is the world-sheet metric. The
interaction term associated with the antisymmetric vector field is
$$ 2iu\epsilon_{\beta \gamma}\partial^\beta a_1
\partial^\gamma a_2  = -i2 \epsilon_{\beta \gamma}a_1 \partial^\gamma a_2
\partial^\beta u = i\epsilon_{\beta \gamma}\epsilon_{jk} a_k\partial^\gamma
a_k \partial^\beta u\eqn\loop$$
The above couplings differ only because one has $\eta_{\beta\gamma}$
and the other has $i\epsilon_{\beta\gamma}$. They
 give opposite contributions to the one loop diagrams since
$${\eta^{\alpha}}_\beta\eta_{\alpha\gamma} = \eta_{\beta\gamma}\qquad
i^2{\epsilon^\alpha}_\beta \epsilon_{\alpha\gamma} = - \eta_{\beta\gamma}
 \eqn\coupling $$
This is exactly what happens in [\Witten, \Nappi].
 Moreover
there are no higher loop contributions to the beta function. Indeed
one can check that no
higher loop diagrams can even be drawn, because all of the interaction
vertices have an external $u$, but the $uu$ propagator vanishes.
Therefore $c=4$ identically in perturbation theory.

The  statement
can be confirmed nonperturbatively by generalizing
the Sugawara construction to non-semi-simple algebras. One way is to
follow the approach in [\Halpern] and check that the following
 tensor
$$L^{AB}={1\over 2}\pmatrix{1&0&0&0\cr
0&1&0&0\cr
0&0&0&1\cr
0&0&1&1-b\cr}\eqn\tensor$$
satisfies the Virasoro master equation
$$L^{AB} = 2L^{AC}\Omega_{CD}L^{DB} - L^{CD}L^{EF}f^A_{CE}f^B_{DF}
- L^{CD}f^F_{CE}f^{[A}_{DF}L^{B]E} \eqn\master$$
This means that one can carry out the Sugawara construction and that
the central charge is $c=2\Omega_{AB}L^{AB}=4$.  A similar
analysis has been carried out for a larger class of non-semi-simple
groups [\olive].

The metric \metric\ can be turned into a more familiar form
by using the following representation for $g$
$$g= e^{a P_1} e^{uJ} e^{fP_1 + vT} \eqn\abel$$
which exhibits the three commuting symmetries of the model
(the left and right action of $P_1$ and the left or right action of
the central object $T$ all commute).
This representation \abel\ can be transformed into
the form \group\ used earlier via
$$a_1\rightarrow a + f \cos u\quad a_2\rightarrow f\sin u
\quad v\rightarrow v + {1\over 2}af\sin u \quad
u\rightarrow u \eqn\change$$
Then the metric \metric\ turns into
$$ ds^2 = da^2 + df^2 + 2 dadf\cos u + 2 dudv + bdu^2\eqn\last$$

The above metric exhibits three commuting symmetries
which are realized as translations
$$ a\rightarrow a + c_1 \qquad f\rightarrow f +c_2
\qquad v\rightarrow v+c_3\eqn\llast$$
for arbitrary constants $c_1, c_2, c_3$. Obviously, translation in
$v$ is a null
vector. The above metric can be recognized as the  metric [\Rosen]
of a plane wave.  This plane wave is monochromatic;
it is an extremely strong wave,
since there are actually in this coordinate system
two singularities in each period -- when $\cos u=\pm 1$.

Of course, the model really has more symmetry
than is manifest in either of the two ways of writing the metric  given
above.  From its origin as a WZW theory, the model clearly describes
a homogeneous space-time; the left and right action of $R$ on itself
gives {\it in toto} a seven dimensional symmetry group of the space-time.
($R$ is four dimensional, but as the left and right actions of the
central generator $T$ coincide, the total number of symmetries coming from the
left and right action of $R$ on itself is seven rather than eight.)
The extra symmetries determine the particular strong field nature of
this plane wave.

Our explicit argument showing that the higher order corrections to the
beta function vanish because there simply are not any suitable Feynman
diagrams is probably relevant to a larger class of plane waves.
\refout
 \end